# Transfeminist AI Governance

Blair Attard-Frost
University of Toronto
Faculty of Information
blair@blairaf.com

March 20, 2025

**Abstract**

This article re-imagines the governance of artificial intelligence (AI) through a transfeminist lens, focusing on challenges of power, participation, and injustice, and on opportunities for advancing equity, community-based resistance, and transformative change. AI governance is a field of research and practice seeking to maximize benefits and minimize harms caused by AI systems. Unfortunately, AI governance practices are frequently ineffective at preventing AI systems from harming people and the environment, with historically marginalized groups such as trans people being particularly vulnerable to harm. Building upon trans and feminist theories of ethics, I introduce an approach to transfeminist AI governance. Applying a transfeminist lens in combination with a critical self-reflexivity methodology, I retroactively reinterpret findings from three empirical studies of AI governance practices in Canada and globally. In three reflections on my findings, I show that large-scale AI governance systems structurally prioritize the needs of industry over marginalized communities. As a result, AI governance is limited by power imbalances and exclusionary norms. This research shows that re-grounding AI governance in transfeminist ethical principles can support AI governance researchers, practitioners, and organizers in addressing those limitations.

## 1. Introduction

Efforts to govern artificial intelligence (AI) are frequently failing to prevent AI systems from causing harm. The harms caused by AI systems have become so significant in scale, scope, and frequency that databases have been created to take inventory of real-world harms (AIAAIC, 2025; AI Incident Database, 2025a). As of March 2025, the AIAAIC Repository contains records of 1905 incidents of AI systems causing harm to individuals and groups, while the AI Incident Database contains records of 931 incidents. The AI Incident Database categorizes AI harms into different levels of severity and types of harm, including physical, psychological, social, political, and economic harms (AI Incident Database, 2025b). The incidents documented in both the AI Incident Database and the AIAAIC Repository implicate a range of AI use cases in these harms. Discriminatory uses of automated decision-making and facial recognition, unsafe autonomous vehicles, as well as fraudulent and hateful AI-generated content are types of AI applications that cause harm with particularly high frequency.

In addition to directly harming people, environmental harms caused by AI systems are also well-documented. The processes required to develop large-scale machine learning models and operate their computing infrastructure are energy- and water-intensive (Li et al., 2023; Luccioni et al., 2024). As a result, the environmental impacts of AI systems have recently become an area of significant concern to researchers, practitioners, policymakers, and communities where the local environment has been damaged by AI development projects (Adarlo, 2023; GPAI, 2021; Lehuedé, 2024; OECD, 2022; Ren and Wierman, 2024; Tessono, 2024).

Although the existence of societal and environmental harms caused by AI systems is clear, the reasons why AI governance practices are so frequently ineffective at preventing those harms are relatively unclear. In the 2010s, AI governance emerged as a field of research and practice to address the social impacts of machine learning technologies. Like AI itself, AI governance has no universally agreed upon definition, but AI governance is often characterized as a system of practices intended to maximize benefits and minimize harms caused by AI systems (Dwivedi et al., 2021; Mäntymäki et al., 2022; Nitzberg and Zysman, 2022). AI governance practices aim to intervene in the impacts of AI systems through the design and implementation of several types of initiatives, such as strategies, policies, government programs, standards, and ethics codes (Birkstedt et al., 2023). AI governance systems have been studied across various contexts and at various scales of activity, including international AI governance activities (Erdélyi and Goldsmith, 2022; Roberts et al., 2023; Tallberg et al., 2023), national AI governance activities (Djeffal et al., 2022; Radu, 2021; Wilson, 2022), local and municipal AI governance activities (Kinder et al., 2023; Wan and Sieber, 2023), sectoral AI governance activities (Kuziemski and Misuraca, 2020; Zuiderwijk et al., 2021), and organizational AI governance activities (Cihon et al., 2021; Mäntymäki et al., 2022). Unfortunately, AI governance practices across these contexts are frequently failing to protect people and the planet.

In this article, I clarify why AI governance is frequently ineffective at preventing AI systems from harming society and the environment: AI governance systems are structurally constrained by power imbalances and exclusionary norms that emerge from relationships between dominant state and industry actors. Drawing upon findings from three empirical studies of AI governance systems in Canada and globally, I apply trans and feminist theories of ethics to (1) focus analysis upon challenges of power, participation, and injustice in AI governance, and (2) re-imagine AI governance as a field of research and practice grounded in transfeminist ethical principles, such as transformative change, fluidity, community, and resistance.

In the next section, I review historical context, epistemologies, and ethical principles that are foundational to a transfeminist approach to AI governance. In Section 3, I describe the research design and methodology that guided my three studies. In Section 4, I summarize key findings and recommendations from those studies. In Section 5, I apply methods of critical self-reflexivity to reflect on the implications of my studies for the future of AI governance. In three reflections on my research, I show that AI governance is often limited by harmful power relations. Based on my reflections, I argue that AI governance must be re-imagined through a transfeminist lens that affords a sharper focus on power imbalances and exclusionary norms, collective resistance, and community empowerment. I conclude in Section 6 with some brief remarks on the need for AI governance researchers and practitioners to address a globally rising tide of anti-trans legislation, digital repression, and administrative violence.

## 2. Transfeminist Ethics & Governance

Transfeminist AI governance is grounded in trans and feminist theories of ethics. *Feminist ethics* describes a diverse set of theories of applied and normative ethics that share a common orientation toward resisting the devaluation of femininity and embracing historically feminized values such as care, relationality, interdependence, and maintenance (de la Bellacasa, 2017; Keller and Kittay, 2017; Mattern, 2018; Moriggi et al., 2020). Shade (2023) observes that a wide range of feminist ethical principles have shaped the evolution of early 2000s media reform movements into the intersectional digital and data justice movements of the late 2010s and 2020s. As part of that evolution, *feminist AI* has emerged as an approach to AI research and practice that applies feminist ethical principles to the development and use of AI systems. Proponents of feminist AI recognize that AI systems are often developed and used under oppressive social conditions. Proponents therefore advocate for enacting feminist principles in AI systems to prevent further harms and promote more beneficial outcomes for oppressed and vulnerable groups (A+ Alliance, 2021; Toupin, 2024; Varon and Peña, 2021; Wellner and Rothman, 2020).

Feminist AI shares much in common with other recent applications of feminist ethics to socio-technical practices, particularly within the area of feminist data ethics and data justice (D'Ignazio and Klein, 2020; Gray and Witt, 2021; Garcia et al., 2020; Marčetić and Nolin, 2022). The ethics of "data feminism" as described by D'Ignazio and Klein is especially notable for applying feminist ethical principles to scrutinize power asymmetries in data science and in AI technologies. Data feminism advances an approach to ethics that is grounded in resistance to inequitable power structures, and in knowledges that are pluralistic, embodied, and contextual. Data feminism requires socio-technical practices that call attention to labor and challenge oppressive AI systems.

Feminist approaches to ethics entail ontological, epistemological, and political assumptions about how and why we should acquire scientific knowledge of reality. As imagined by feminist scholars such as Haraway (1988), Harding (1992, 1995), and Suchman (2007), *feminist standpoint epistemologies* understand scientific knowledge as situated in social and historical contexts. Knowledge is embodied in social agents with partial perspectives of their world. Knowledge is produced pluralistically across many communities with differing norms, values, beliefs, abilities, and lived experiences.

Feminist standpoint epistemologies contrast with traditional empiricist and positivist epistemologies. Traditionally, positivism understands scientific knowledge as disembodied and disembedded from social life, value-neutral, and politically impartial. Positivist knowledge is produced by individual scientists in accordance with a universally shared standard of rigor and objectivity. Harding (1992) explains that feminist scientific standpoints have the potential to offer greater rigor and objectivity than traditional positivist standpoints by "critically identifying all of those broad, historical social desires, interests, and values that have shaped the agendas, contents, and results of the sciences much as they shape the rest of human affairs" (p. 359). A feminist standpoint can thereby privilege historically marginalized perspectives to produce knowledge that "can be for marginalized people . . . rather than for the use only of dominant groups in their projects of administering and managing the lives of marginalized people" (p. 445). For many Black and intersectional feminists, standpoint epistemologies provide ground for producing collective knowledge and building epistemic power from marginalized experiences across intersections of race, gender, and class (Collins, 2000; Crenshaw, 1991; Dotson, 2014).

Researchers have recently applied feminist epistemologies to study AI systems that harm marginalized groups, and to recommend actions for empowering those groups with greater agency in the design and governance of those systems (Benjamin, 2019; Birhane, 2021; Hancox-Li and Kumar, 2021; Kong, 2022; McQuillan, 2022; Ricaurte, 2022; Schelenz, 2022; Widder, 2024). These studies demonstrate that feminist standpoints are effective for analyzing the social and historical contexts of AI governance activities. Feminist standpoints enable us to identify how power relations in AI governance activities can be changed to prevent harms to marginalized groups.

Trans researchers and activists extend feminist theories of ontology, epistemology, and ethics into *transfeminist ethics*. Transfeminist ethics refers to theories of applied ethics that are grounded in the lived experiences of trans women, trans femmes, and other transgender and gender-diverse people who self-identify with the umbrella term "trans." Transfeminist ethics therefore centers ethical principles that are of particular importance to trans experiences of existing within cis-hetero-normative social systems. These principles include transformative change, anti-normativity, fluidity, agency, community, care, and resistance against governance systems that perpetuate anti-trans violence (Barad, 2015; Galpin et al., 2023; Malatino, 2020; Marvin, 2019; McFadden et al., 2024; Spade, 2015; van der Drift and Raha, 2020).

In their analysis of the political implications of transfeminist principles, van der Drift and Raha (2020) argue for adopting transfeminist practices of ethics and governance that are "active and anti-normative, rather than defined in a stable form," recognizing that "the dynamism of the term 'trans' indicates that we must attend to questions of agency and structures of action" (p. 13). Transfeminist ethics therefore demands not just theoretical commitments, but also practical commitments to particular actions and political goals. Transfeminist ethics seeks to protect the fluidity and vitality of trans lives against various forms of anti-trans violence, including physical, psychological, social, economic, and administrative violence. Transfeminist ethics seeks to cultivate collective resistance against dominant social structures that cause harm to trans communities, and to radically change those harmful structures. Transfeminist ethics seeks to secure greater agency and power for trans communities and other marginalized communities.

Accordingly, trans practices of governance are grounded in community-led design and collective organizing against governance systems that cause harm to trans people and other marginalized groups (Costanza-Chock, 2020; Nownes, 2019; Verloo and van der Vleuten, 2020). Trans governance involves polycentric capacity-building, resource-sharing, and mutual care arrangements within localized networks of trans people. These practices enable trans communities to co-create rules and norms with one another and with sympathetic institutions, thereby transforming narrow relations of trans political inclusion into stronger relations of accountability and justice (Davidson, 2007; Malatino, 2020, 2022; McFadden et al., 2004; van der Drift and Raha, 2020; Verloo and van der Vleuten, 2020).

In my research, I apply transfeminist ethical principles and practices through an approach I refer to as *transfeminist AI governance*. I define transfeminist AI governance as *the application of transfeminist ethical principles to the study and practice of AI governance*. Researchers have found that AI systems put trans people at risk of experiencing many types of physical, psychological, social, and economic harms. These include harms caused by trans-exclusionary healthcare automation, recruitment and hiring automation, facial and gender recognition, security and law enforcement applications, and generative AI applications (Costanza-Chock, 2018; Keyes, 2018; Scheuerman et al., 2021; Scheuerman et al., 2019; Ungless et al., 2023). However, rather than studying the ethical issues arising from those and other specific anti-trans use cases of AI, I apply principles and practices of transfeminist ethics to study AI systems and their governance more generally. My approach to transfeminist AI governance centers a broad scope of ethical issues and governance interventions, all of which are of importance to the protection and flourishing of trans and other marginalized communities impacted by AI systems. There are six main ethical issues and governance interventions I focus on in my research. These six focal points collectively constitute a *transfeminist lens*:

(1) *Socio-political & socio-economic conditions*: Conditions of political and economic inclusion, exclusion, injustice, and accountability in AI systems and their governance.

(2) *Inequitable value co-creation*: The co-creation of inequitable AI systems and governance systems through networks of dominant actors, marginalized actors, resource dependencies, and institutional arrangements.

(3) *Norms & values of dominant actors*: Exclusionary norms and values underlying state-led and industry-led AI governance systems.

(4) *Harms of dominant actors*: Harmful outcomes of AI governance actions and inactions taken by dominant state actors and industry actors.

(5) *Opportunities for top-down change*: Practical opportunities for preventing harm and securing greater power for marginalized communities by changing state-led and industry-led AI governance systems.

(6) *Opportunities for bottom-up change*: Practical opportunities for preventing harm and securing greater power for marginalized communities by developing AI governance systems led by communities, workers, and civil society organizations.

As we will see in Sections 4 and 5, AI systems have a multitude of impacts on society and the environment. Because of the global scale, high frequency, and complexity of those impacts, the field of AI governance covers such a vast space of ethical issues and governance interventions that it is impossible to comprehensively cover every conceivable issue and intervention within a scope of a single study. Transfeminist AI governance therefore provides a meta-theoretical heuristic for orienting analysis of AI systems and AI governance systems toward the focal points outlined above.

## 3. Research Design & Methodology

To investigate ethical issues and practical challenges facing AI governance, I collaborated with three co-authors to conduct three empirical studies with three corresponding research objectives (see Table 1). The integrative review and semi-systematic review presented in Studies 1 and 2 were conducted in accordance with Snyder's (2019) guidelines for literature review methodology. Study 2 applied content analysis methods (Hsieh and Shannon, 2005; Nowell et al., 2017) to analyze government documentation, identifying challenges and opportunities for strengthening AI governance initiatives in Canada and in other jurisdictions facing similar challenges. Study 3 combined methods of content analysis and service systems analysis (Frost et al., 2019) to collect and analyze data from 20 interviews with government leaders and subject matter experts about Canada's national AI governance system.

*Table 1: Research design features of three studies of AI governance alongside each study's corresponding research objective.*

| Subject of study & citation | Research objectives | Research design features |
|---|---|---|
| **Study 1:** Global AI Value Chains (Attard-Frost and Widder, 2024) | *AI Impacts*: Determine what types of benefits and harms AI systems are capable of causing, the actors responsible for those benefits and harms, the actors impacted by those benefits and harms, and the activities through which those benefits and harms are caused. | • Development of a theoretical framework detailing the process through which benefit and harm are co-created across the value chains of AI systems.<br>• Application of the framework to an integrative review of 67 sources on ethical and governance issues in global AI value chains. |
| **Study 2:** AI Governance Initiatives in Canada (Attard-Frost, Brandusescu, & Lyons, 2024) | *AI Governance Initiatives*: Determine what types of AI governance initiatives have been created to intervene in those impacts, and the extent to which those initiatives are effective or ineffective at intervening in those impacts. | • Semi-systematic review of documentation from 84 federal and provincial AI governance initiatives in Canada.<br>• Content analysis of the documentation to identify themes, challenges, and opportunities in Canada's AI governance initiatives. |
| **Study 3:** AI Governance Systems in Canada (Attard-Frost and Lyons, 2024) | *AI Governance Systems*: Determine how those initiatives function as part of larger AI governance systems that exist across multiple contexts and levels of scale. | • Development of a framework for multi-scale analysis of AI governance systems.<br>• Application of the framework to collect and analyze data about Canada's national AI governance system from interviews with 20 government leaders and subject matter experts. |

Following the completion of the three studies, I independently applied a critical reflexivity methodology called the *4Rs* (Abbott et al., 2024) to review findings and reflect upon the research in relation to the six focal points of my transfeminist lens. As described by Abbott et al. in their review of the literature on reflexive methodologies, the 4Rs–*retrospection*, *representation*, *review*, and *reinterpretation*–can be applied by information systems researchers to collectively reflect or self-reflect upon research processes, practices, and findings from previously completed projects. The 4Rs enable researchers to retroactively extend data and analysis from past projects into new findings and discussions. I applied the 4Rs to: (1) engage in *retrospection* about my personal experience of carrying out the three studies, (2) *represent* my retrospection in the form of written reflections on my research, (3) *review* findings from the three studies and synthesize research findings with my written reflections, and (4) *reinterpret* the findings through the analytical lens of transfeminist AI governance.

A clear statement of researcher positionality is essential for rigorous critical self-reflexivity. My prior research and my retroactive application of the 4Rs to reflect and expand upon that research are all shaped by my positionality. I am a white nonbinary trans woman living in Canada who has had the privilege of working on applied research, AI development, and information management projects in public sector, private sector, and academic organizations. Over the course of working and studying in cis-hetero-normative and male-dominated social spaces as a trans femme, I have become greatly concerned with patterns of harm and marginalization that I have witnessed and personally experienced within academic and professional environments. Though I have been subjected to some forms of anti-trans harm in my professional and personal life, my experience living in Canadian society as a white settler has afforded me material and institutional advantages over the course of my life that many Indigenous and racialized persons do not benefit from. Knowing that Canadian society and Canada's governance systems have historically been structured to uphold those relations of advantage and disadvantage, I endeavor to remain mindful of how those relations might apply to the AI systems and governance systems I study, and to discuss those relations when they appear in my research.

Key findings and recommendations for researchers, practitioners, and policymakers from each of the three studies are reviewed in the following section. Retrospective reinterpretations of the findings and recommendations are then presented in Section 5 in a series of three pieces of reflective writing with three subject headings: (1) *AI governance is an exercise of power*, (2) *Top-down AI governance is limited by power imbalances*, (3) *AI governance must be transformed*.

**4. Findings & Recommendations**

*Table 2: Summary of findings from Studies 1, 2, and 3.*

| Subject of study & citation | Key findings from research |
|---|---|
| **Study 1:** Global AI Value Chains (Attard-Frost and Widder, 2024) | <ul><li>AI systems are capable of causing benefits to efficiency, productivity, and quality of products and services.</li><li>AI systems are capable of causing harm to marginalized populations, political institutions, social trust, public services, and ecosystems.</li><li>Beneficial and harmful impacts of AI systems are caused by developers, users, service providers, data center operators, manufacturers, governments, communities, and workers.</li><li>AI value chains enable co-creation of value through integration of software, hardware, knowledge, finances, and governance resources throughout the lifecycle of AI systems.</li></ul> |
| **Study 2:** AI Governance Initiatives in Canada (Attard-Frost, Brandusescu, & Lyons, 2024) | <ul><li>Policy instruments, programs, strategies, standards, and ethics statements have been used to intervene in the impacts of AI systems.</li><li>In Canada, AI governance initiatives are often effective at prioritizing interventions in Canadian industry and innovation, AI technology production and use, and AI research.</li><li>In Canada, AI governance initiatives are often ineffective at reporting on post-implementation outcomes, securing public trust in AI, intervening in a diversity of AI impacts, and cultivating strong coordination between governments, sectors, and civil society.</li></ul> |
| **Study 3:** AI Governance Systems in Canada (Attard-Frost and Lyons, 2024) | <ul><li>AI governance initiatives function as activities within AI governance systems.</li><li>The activities that occur within AI governance systems are co-created by actors, resources, networks, logics, and rules that interact across multiple contexts and levels of scale.</li><li>AI governance systems and the initiatives that are designed and implemented within them are structured by the values, perceptions, and capacities of many actors.</li></ul> |

From the findings presented in Studies 1, 2, and 3, we contributed a total of 17 recommendations for advancing future AI governance research and practice to researchers, practitioners, and policymakers. In summary, the recommendations are:

**Recommendation 1:** Researchers should conduct more empirical and action research into the specific ethical concerns, value chain actors, and resourcing activities.

**Recommendation 2:** Researchers should develop and apply theories and methods for systematically modeling AI value chains, analyzing a diverse range of ethical concerns in those value chains, and enacting interventions in those value chains.

**Recommendation 3:** Practitioners and policymakers should implement ethical sourcing practices across value chains that provide resource inputs to or receive resource outputs from AI systems.

**Recommendation 4:** Researchers should study the outcomes of Canada's AI governance initiatives.

**Recommendation 5:** Researchers should study challenges to public trust in Canada's AI governance initiatives.

**Recommendation 6:** Researchers should study the effects of prioritizing intervention in some types of AI impacts over others on the outcomes of Canada's AI governance initiatives.

**Recommendation 7:** Policymakers and public servants should specify success measures for initiatives and routinely publish information on the outcomes of Canada's AI governance initiatives.

**Recommendation 8:** Policymakers and public servants should collaborate more directly with the public on designing and implementing Canada's AI governance initiatives.

**Recommendation 9:** Policymakers and public servants should account for a greater variety of AI impacts when designing and implementing Canada's AI governance initiatives.

**Recommendation 10:** Policymakers and public servants should launch a new initiative to cultivate a more unified national approach to AI governance in Canada.

**Recommendation 11:** Researchers should conduct additional analysis of topics in our dataset of perceptions of Canada's national AI governance system.

**Recommendation 12:** Researchers should further investigate institutional and ecosystem phenomena in Canada's national AI governance system.

**Recommendation 13:** Researchers should apply our AI governance systems analysis framework, data, and findings to other AI governance research contexts.

**Recommendation 14:** Practitioners and policymakers should implement new collaboration and coordination mechanisms in Canada's national AI governance system.

**Recommendation 15:** Practitioners and policymakers should create guidance for designing and implementing participatory AI governance initiatives in Canada.

**Recommendation 16:** Practitioners and policymakers should expand access to key resources needed for effective AI governance practices in Canada.

**Recommendation 17:** Practitioners and policymakers should advance diversity, equity, and inclusion in Canada's AI governance activities.

If effectively implemented, these 17 recommendations could support greater prevention of harms caused by AI systems. However, effective implementation of these 17 recommendations will face significant barriers in navigating the power relations inherent to AI systems and AI governance systems. In Studies 1, 2, and 3, we determined that imbalances in political power, economic power, computational power, cultural power, and other forms of power between actors often present barriers to enacting change in AI systems and AI governance systems. Additional barriers to change are imposed by norms of exercising power to organize actors and to provide them with resources to do the work of governance.

*Transformative change*–change in which unjust norms and power structures are dismantled to prevent further harm to historically marginalized groups–is especially difficult to enact in AI systems and AI governance systems. These systems are built atop a stack of other unjust computational, technological, and social systems that long preceded the recent advent of data-intensive machine learning applications (Crawford, 2021; Crawford and Joler; 2023; Crawford and Joler, 2019). How, then, can AI governance possibly be made more effective at preventing harms–and particularly, at preventing harms to marginalized communities–if studies, practices, and systems of AI governance are all so limited by power structures and by histories of domination and oppression?

As I've reflected on the findings and implications of my research, some more precise explanations for the ineffectiveness of AI governance have emerged. I've also reflected on some necessary pathways forward for re-imagining what "AI governance" fundamentally is and what the work of AI governance ought to entail. The following section details those explanations and future pathways through three thematically interlinked reflections on my findings from Studies 1, 2, and 3.

**5. Reflections on AI Governance**

*5.1. Reflection 1: AI governance is an exercise of power*

Studies 1, 2, and 3 all demonstrate that AI governance frequently proves ineffective at preventing AI systems from harming people and the environment because studies, practices, and systems of AI governance are limited by various forms of power relations. These power relations include resource dependencies, rule-making and rule-taking activities, and norms of political and economic inclusion. Power relations are deeply embedded in networks of actors and in the logics, rules, and norms involved in their co-creation of AI governance systems. The effectiveness of an AI governance activity at preventing AI systems from causing harm is therefore derived from the perceptions and powers of the actors included in that activity: AI governance emerges from the underlying values and logics of those actors, the qualities those actors perceive as constituting "effective" vs. "ineffective" AI governance, and the resources and capacities those actors are able to exercise to enact their shared values and logics.

In researching and writing Study 1, I often felt overwhelmed by the immense challenges of context and scale implicated in the societal and environmental impacts of AI systems. The resources required to develop and operate AI systems are distributed across global value chains. This results in complex issues of jurisdiction and transnational impact that AI governance researchers, practitioners, and policymakers must contend with. Industry actors making voluntary commitments to more robust ethical sourcing standards at socially and environmentally impactful points throughout the AI value chain (as suggested in *Recommendation 3*) could help address challenges of transnational context and global scale in principle. In practice, however, effective global implementation of ethical sourcing practices faces political and economic barriers arising from the power relations embedded in AI systems and their governance. Veale, Matus, and Gorwa (2023) highlight several tensions limiting the global implementation of ethical and regulatory standards for AI. Dominant industry actors consolidate expertise within their companies and lobby governments to influence regulation. Industry also outsources labor and AI development tasks to extraterritorial regulatory havens. These outsourcing practices create perverse incentives for obtaining foreign investment, resulting in race-to-the-bottom dynamics in the development of regulations and standards.

Given these tensions of political and economic power in AI governance, voluntary adoption of robust ethical standards for preventing harms to society and the environment will have limited effectiveness. Governments could impose legal and regulatory requirements for industry actors to comply with more robust ethical standards across the AI value chain, but these top-down pushes for comprehensive, compliance-based regulatory frameworks are also limited by power relations. Relations of political and economic power have limited the effectiveness of AI regulation across several jurisdictions. For instance, although the European Union's *AI Act*–a general-purpose, cross-sectoral regulatory framework for AI systems–was developed through extensive multistakeholder consultations, the final text of the EU AI Act contains limited protections for workers and the environment. In addition to those limitations, some have criticized the EU AI Act for containing limited protections for human rights, and for scaling back the compliance requirements and protections set forth in earlier versions of the AI Act in response to pressure from industry stakeholders and lobbyists (Access Now, 2024; Amnesty International, 2023; Corporate Europe Observatory, 2023).

Similar power dynamics can be observed in the United States, where there is no general-purpose regulatory initiative comparable to the EU's AI Act, but rather, a patchwork of executive, administrative, and legislative initiatives at multiple levels of government and with varying degrees of legal force. Some voluntary instruments such as the Biden White House's *Blueprint for an AI Bill of Rights* (The White House, 2022) and the National Institute of Standards and Technology's *AI Risk Management Framework* (NIST, 2023) were created in consultation with several public sector, private sector, and civil society organizations. However, the US AI industry exercises considerable political, economic, and cultural power across the nation's AI governance system. Companies that dominate the AI industry–such as Google, Amazon, Microsoft, and OpenAI–shape US AI governance to suit their strategic priorities and economic interests. Industry power is exercised through regular engagement with government officials, extensive participation in the development of policy and regulation, and privileged access to multistakeholder meetings organized by Congressional and White House committees (Henshall, 2024; Hine, 2024; Johnson, 2023; Merica, 2024; The White House, 2023). Simultaneously, the US AI industry works to normalize futurological narratives about AI's supposedly utopian promise and apocalyptic peril–and the ability of AI developers to navigate society away from peril and toward greater promise–as a cultural pretext for policymaking and for media coverage of AI governance activities (Altman et al., 2023; Barakat, 2024; Dandurand et al., 2023; Gebru and Torres, 2024; Herrman, 2024; OpenAI, 2023). Executive action taken by the second Trump administration rolls back AI policy and regulatory mechanisms established under the Biden administration, further fortifying the already immense power of the US AI industry (The White House, 2025).

In researching and writing Study 2, I became more aware of the extent to which industry power also dominates the AI governance priorities of Canada. This in itself is not a new finding: in her landmark empirical study of public and private investment activity in Canada's AI ecosystem, Ana Brandusescu (2021) demonstrates clearly that "concentrations of power provide advantages to a handful of entities with financial resources, data, and technologies across a few universities and affiliated research nonprofits, startups, and international (big) tech companies" (p. 7). The findings we present in Study 2 validate and build upon her earlier findings regarding power imbalances across Canada's AI governance initiatives.

Study 2 demonstrates that Canadian federal and provincial AI governance initiatives launched from 2017 to 2022 prioritize the development of Canadian industry, technological innovation, and economic growth over protections for vulnerable social groups, workers, and the environment. Even policy instruments created for the ostensible purpose of preventing harms caused by AI systems–such as the *Artificial Intelligence and Data Act* (AIDA) and the *Directive on Automated Decision-making*–have substantial gaps in the scope of harms they cover and in their enforceability (Attard-Frost, 2022; Attard-Frost, 2023a; Brandusescu and Sieber, 2022; Scassa, 2021; Scassa, 2023; Tessono et al., 2022). The AIDA in particular was widely criticized for many reasons. The AIDA failed to meaningfully include civil society and marginalized communities in its policy development process. In addition, the AIDA did not include a wide range of collective harms and environmental harms in its scope, and it did not guarantee adequate protections against harms caused to vulnerable groups caused by AI systems (The Dais and Centre for Media, Technology and Democracy, 2023).

Some more recent Canadian AI governance initiatives address a greater scope of societal and environmental impacts. For example, the *Voluntary Code of Conduct on the Responsible Development and Management of Advanced Generative AI Systems* was published by Innovation, Science and Economic Development Canada (ISED) in September 2023 (Innovation, Science and Economic Development Canada, 2023a). The Code of Conduct is intended to secure voluntary commitments to a set of ethical and safety standards from developers, deployers, and operators of generative AI systems. Also in September 2023, the Treasury Board of Canada Secretariat (TBS) published their *Guide on the use of Generative AI* (Government of Canada, 2024). The TBS Guide provides federal public servants with voluntary guidance for ensuring quality, protection of information, and human autonomy in their use of generative AI systems. In addition to those voluntary instruments, amendments were proposed to the AIDA by the Minister of ISED in November 2023 to address gaps in the legislation (Minister of Innovation, Science and Industry, 2023). The amended version of the AIDA added specificity to the legislation's definitions, clarified regulator and stakeholder responsibilities across the AI value chain, and moved the legislation into closer alignment with international frameworks such as the EU's AI Act.

As I reflected on these more recent initiatives and amendments to the AIDA, I noticed that although their scope represented a more clear and wide range of AI impacts than earlier initiatives, many underlying challenges of industry prioritization, public participation, and public accountability still remained. The ISED Code of Conduct was criticized for an extremely short consultation period of less than one month, for being developed out of invitation-only meetings with a limited group of stakeholders, and for limitations in its scope, accountability measures, and enforceability (Attard-Frost, 2023b; Innovation, Science and Economic Development Canada, 2023b; Karadeglija, 2023). The ISED Code of Conduct and the TBS Guide both continue to contribute to a lack of transparency and accountability in the outcomes of Canadian AI governance initiatives.

The TBS Guide received some praise for its ambitious scope, and 46 organizations have voluntarily signed onto the ISED Code of Conduct as of March 2025. However, it is not clear if these initiatives have had any significant effect on the operations of Canadian AI developers and users. There has been no formalized post-implementation monitoring, reporting, or evaluation of the effectiveness of these guidance documents at achieving their goals of risk mitigation and harm prevention.

Meanwhile, the Minister of ISED's proposed amendments to the AIDA were met with mixed reception: while some changes to the legislation were praised, there were also concerns that the amendments still did not sufficiently address criticisms of the legislation's gaps in scope, specificity, enforcement powers, and public consultation (Canadian Association of Professional Employees, 2024; Canadian Chamber of Commerce, 2024; OpenMedia et al., 2024). Some critics observed that the consultation activities that did occur greatly overrepresented industry actors and their interests (Castaldo, 2023; Clement, 2023). In their written brief to the parliamentary committee studying the legislation, the Assembly of First Nations (2023) announced that they may take legal action against the government for failing to uphold a constitutional responsibility to consult with Indigenous peoples during the drafting of the legislation.

As the parliamentary process of studying and amending the AIDA proceeded into 2024, the federal government's 2024 budget announcement further underscored the government's priorities: $2 billion was to be invested in expanding computing infrastructure for Canadian industry and AI researchers, $200 million was allocated to commercializing AI technologies and accelerating AI adoption across critical sectors, while only $50 million was allocated to "supporting workers who may be impacted by AI, such as creative industries" (Prime Minister of Canada, 2024). The proroguing of Parliament in January 2025 has cast some uncertainty upon the future of AI legislation and governance in Canada. However, these more recent developments confirmed to me that our findings in Study 2 from reviewing 84 initiatives launched from 2017 to 2022 have remained valid. Canadian AI governance still prioritizes the needs of industry over broader societal needs. Canadian AI governance still fails to ensure meaningful participation from and accountability to marginalized communities and the broader public. Canadian AI governance is still ineffective at preventing AI systems from causing harms to society and the environment. In Canada, support for and development of industry power is treated with greater importance than public empowerment.

*5.2. Reflection 2: Top-down AI governance is limited by power imbalances*

My experience witnessing repeated shortcomings of government accountability, transparency, and public participation across so many Canadian AI governance initiatives made me increasingly concerned that the Canadian nation-state may be too structurally limited to effectively prevent AI systems from harming society and the environment. My concern was compounded in researching and writing about Canada's national AI governance system for Study 3. In Study 3, I conducted interviews with 20 government leaders and subject matter experts throughout 2023. These interviews further clarified for me that Canada's industry-first approach to AI governance is in large part a product of the political, economic, and cultural power that industry exercises over government.

During these interviews, the logics, limitations, and rules that participants described to me made it strikingly clear that Canada's national AI governance system is shaped and constrained by imbalanced power relations. The logics for engaging in AI governance activities described most frequently by participants indicated that dominant industry actors are able to strongly steer the direction of Canadian AI governance by exercising their political and economic power. *Maximize profit & shareholder value from AI adoption* (a topic described in 6 interviews) indicates that AI governance is often practiced in Canada with the goal of securing financial benefits and greater economic power for private interests. *Achieve balance between public & private interests in AI outcomes* (a topic described in 6 interviews) indicates that AI governance practices intended to secure more beneficial AI adoption outcomes for the public are somehow reconcilable with the goal of securing financial benefits and greater economic power for private interests. This balancing of public and private interests functions through a variety of other logics perceived by the participants, most notably, *strengthen public/consumer trust in AI applications* (described in 4 interviews) and *ensure AI development & use contributes to economic development* (3 interviews). The rationale implied by these logics is highly speculative: by stimulating AI adoption and ensuring AI applications are trusted by Canadian consumers, the development and use of those trustworthy AI applications will contribute to Canada's economic development, which will in turn contribute to economic benefits being accrued by and fairly balanced between private stakeholders and the general public. In the process of analyzing these logics and writing Study 3, I reflected on a part of our analysis from Study 2:

> In Canada's stimulation approach to AI governance, societal benefit is typically assumed to be a secondary epiphenomenon of economic benefits accrued through technological innovation. There is no empirical evidence to support such assumptions about the socio-economic impacts of AI systems. In fact, many studies of the political economy of AI indicate that without broad-based and cross-cutting interventions in industries, technologies, societies, workforces, and digital infrastructures, a stimulation approach to AI governance might instead cause negative societal impacts. Without adequate counterbalances, expanding industry's capacity to develop and use AI systems may compound existing concentrations of capital, technology, data, and other resources in a small handful of dominant industry actors (Attard-Frost, Brandusescu, & Lyons, 2024, p. 10).

This empirically tenuous *rising tide lifts all boats* rationale is conditional on the success of a long series of complex, well-coordinated policy interventions intended to counterbalance the domineering power of industry. Despite the many speculative leaps this rationale requires, it is so deeply embedded into the logics of Canada's national AI governance system that it functions as a cultural narrative: it is an implicit pretext for engaging in AI governance activities that is observable across multiple studies. This cultural narrative of broad-based societal benefit is made durable and enduring by strong support and repeated deployment of the narrative from private sector and public sector organizations. Other logics that were frequently perceived by participants in the research conducted for Study 3 also function in service to this cultural narrative, such as *align AI systems with organizational values* (a topic described in 5 interviews), *cultivate a trustworthy brand image* (3 interviews), *enable greater access to international markets* (3 interviews), and *create AI systems that are interoperable across markets & jurisdictions* (3 interviews). Values-based AI adoption, brand trustworthiness, interoperability, and market access do not primarily serve the public interest: these logics for engaging in AI governance primarily serve the AI adoption goals of Canadian industry and their private shareholders. These logics are premised on a speculative narrative that a balance of public and private benefits will somehow be achieved as a downstream effect of well-calibrated organizational values, trustworthy brands, and interoperable AI markets.

In conducting research interviews for Study 3, the limitations and rules described by participants further concerned me, as they signaled that Canada's national AI governance system has been structured by powerful norms that may already be too solidified to be susceptible to transformative change. Limitations of *AI governance knowledge and expertise* and *financial resources* within organizations are widespread (topics described in 10 and 5 interviews), making it difficult for smaller private sector and civil society organizations to effectively practice and meaningfully participate in AI governance. AI policy development processes are strongly influenced by the norms and cultures of organizations (a topic described in 7 interviews). This means that if a small handful of powerful public sector organizations co-create AI policy with a small handful of powerful private sector organizations, a result will be that regular involvement of private interests in public policy processes becomes a cultural norm. This emerging norm of private control over public AI policy is reflected in the domain of regulatory development, where the *privatization of audit, compliance, & regulatory services* (described in 4 interviews) is a contentious issue. The normalization of dominant industry actors steering the politics and economics of Canada's national AI governance system also reinforces existing *norms of & imbalances in political & economic power* (4 interviews), contributing to an *emerging norm of accountability & enforceability gaps* (3 interviews). With gaps in accountability and enforceability, the public sector becomes so dependent on specialized AI governance knowledge, expertise, and other resources provided by the private sector that it is normal for the public sector to design AI policy instruments with light-touch accountability requirements and weak enforcement capabilities. If Canada's national AI governance system so greatly prioritizes the regulatory needs and perspectives of dominant industry actors, then it will be difficult to effectively prevent AI systems from causing harm to marginalized communities and society more broadly.

Many of the government leaders and subject matter experts I interviewed were optimistic that power imbalances in Canada's national AI governance system can be counterbalanced through top-down government intervention. I am less optimistic. Participants suggested opportunities for improvement such as *create more opportunities for public participation in AI governance* (7 interviews), *more diversity in AI governance activities* (6 interviews), *implement stronger participatory design & governance practices* (6 interviews)*,* and *stronger public accountability & oversight in governance activities* (4 interviews). However, enacting these opportunities at a national scale will require a protracted and concerted effort from public sector, private sector, and civil society organizations. Working together closely, these organizations will need to build greater power and resources for marginalized groups, along with greater capacities for participatory governance, co-regulation by a diversity of actors, proactive accountability measures, and robust protections against harms caused by AI systems. Unfortunately, this kind of diverse, equity-seeking approach to collaboration seems highly unrealistic. In the case of powerful government institutions and industry actors that benefit from power imbalances in Canada's national AI governance system, this would in many cases mean acting against their own established norms and their own perceived interests.

In addition to conflicting interests and political challenges, a national scale effort to diminish power imbalances in AI governance would encounter practical challenges navigating gaps in coordination and collaboration discussed in Studies 2 and 3. The networks needed to support broadly inclusive and equitable collaboration across governments, sectors, and civil society are not well developed, making *Recommendations 8* and *17* difficult to implement effectively (*Recommendation 8:* collaborate more directly with the public on designing and implementing Canada's AI governance initiatives, and *17*: advance diversity, equity, and inclusion in Canada's AI governance activities). In principle, implementation of *Recommendations 10* and *14* (*Recommendation 10:* implement new collaboration and coordination mechanisms, and *14*: launch a new initiative to cultivate a more unified national approach to AI governance) would further develop these networks, creating the infrastructure needed to enable more participatory forms of governance. In practice, however, implementing these recommendations would be a resource-intensive endeavor, again requiring dominant actors to act against their own established norms and perceived interests by empowering actors at the margins of Canada's national AI governance system with greater agency to change the system's structures and operations.

Without clear incentives for dominant actors to empower civil society and marginalized communities, other recommendations directed at public servants and policymakers are also likely to have limited effectiveness. The effective implementation of *Recommendations 15* and *16* (*Recommendation 15:* create guidance for participatory AI governance initiatives, and *16:* expand access to key resources needed for effective AI governance practices) are dependent on powerful actors with existing networks of knowledge, financial, computational, legal, and policy resources voluntarily redistributing their resources to relatively less powerful actors. The effective implementation of *Recommendation 7* (specify success measures for initiatives and routinely publish information on the outcomes of initiatives) is dependent on government institutions and individual public managers voluntarily subjecting their operations to greater public scrutiny. This recommendation also depends on civil society organizations possessing the resources and capacities needed to hold government institutions accountable when they are not transparent about success measures and outcomes. The effective implementation of *Recommendation 9* (account for a greater variety of AI impacts in governance initiatives) is dependent on government institutions going against established political, cultural, and epistemic norms of prioritizing the impacts of AI systems on industry and technological innovation over other societal and environmental impacts.

As I continued reflecting on the embedding of these power imbalances and resource dependencies in Canada's national AI governance system, I recognized that *top-down AI governance*–a system of governance in which marginalized communities are dependent upon resources and leadership provided to them by brokers of state power and industry power–is not sufficient for enacting a transfeminist approach to AI governance. To enact transfeminist principles of transformative change, agency, fluidity, community, and resistance in AI governance, systems of *bottom-up AI governance* are necessary. While top-down AI governance is set in motion by alliances of state power and industry power, bottom-up AI governance is set in motion by the power of collective action at smaller scales of organizational activity, such as communities and workplaces.

AI governance activity already occurs regularly at smaller scales. A recent flashpoint in bottom-up AI governance can be seen in the resistance of creative communities and workers against harmful generative AI systems. Popular text-generating, image-generating, and audio-generating AI applications such as ChatGPT, DALL-E, Stable Diffusion, Midjourney, and MuseNet are trained on millions of copyrighted works that have been scraped from the web and used for model training without consent from or compensation to the creators of those works (Jiang et al., 2023). In response, online communities and social movements such as #CreateDontScrape (Create Don't Scrape, 2024) and labour unions such as the Writers Guild of America (WGA) and Screen Actors Guild - American Federation of Television and Radio Artists (SAG-AFTRA) have collectively organized against developers and operators of generative AI systems. The 2023 WGA strike resulted in the union establishing regulations for the training of generative AI models on union-protected materials, and for allowing workers to decide if and how generative AI should be used in their own work processes (Writers Guild of America, 2023). The 2023 SAG-AFTRA strike resulted in regulations being established for the development and use of synthetic digital replicas of performers (Patten, 2023). These community-led and worker-led initiatives offer a powerful lesson for the future of AI governance: if top-down AI governance systems are structurally limited by power imbalances, bottom-up AI governance can provide communities and workers a viable alternative to dependence on resources and regulatory leadership from state and industry actors.

*5.2. Reflection 3: AI governance must be transformed*

AI governance research, practices, and systems must be radically transformed to more effectively prevent harm to marginalized communities. In her analysis of state-led and industry-led initiatives for governing generative AI systems, Inga Ulnicane (2024) coins the term "governance fix" to describe highly centralized, top-down initiatives that enact a "narrow and technocratic approach" to AI governance (p. 1). Instead of governance fixes that reinforce existing power imbalances, Ulnicane recommends enacting more pluralistic, polycentric, and participatory approaches to co-governance that seek to transformatively change structural imbalances of power. Mere governance fixes–new initiatives, policy amendments, and minor operational changes within the narrow perimeters of existing power structures–are insufficient without a deeper re-imagining and reconfiguration of the structures, norms, and possibilities of AI governance. By re-imagining AI governance through a transfeminist lens, we can envision alternative possibilities for AI governance that are less harmful and more effective, with effectiveness understood in relation to transfeminist principles of transformative change, agency, fluidity, and community.

I outline here a preliminary agenda for a transfeminist approach to AI governance that aims to transformatively change future AI governance research, practices, and systems. This agenda contains three high-level actions for researchers, practitioners, policymakers, community organizers, advocates, and other stakeholders to implement in parallel: (1) Expand the scope of AI and AI governance, (2) Oppose AI governance systems that exclude marginalized communities, (3) Co-create bottom-up AI governance systems to reduce dependency on industry and the state.

**Action 1: Expand the scope of AI and AI governance.** Studies 1, 2, and 3 all show that "AI" is an ambiguous phenomenon. The meaning of AI–once a purely theoretical concern for an esoteric enclave of computer scientists, philosophers, and futurists–has in recent years become a matter of enormous practical and political consequence. In her 2021 book *Atlas of AI*, Kate Crawford observes that "each way of defining artificial intelligence is doing work, setting a frame for how it will be understood, measured, valued, and governed" (p. 7). The ways in which the term AI is interpreted has a cascading effect on how AI governance is studied, practiced, and systematized. The ontological boundaries of AI that are encoded into now-emerging AI laws, policy instruments, and standards will cascade into the system design requirements of AI developers and operators. Ontologies of AI also cascade into business rationales for public and private sector AI investment, procurement, and management decisions. Theoretical differences in how to understand and interpret the meaning of AI will ultimately result in practical differences: AI systems will be developed, used, and governed differently based on what those systems are understood to consist of. If AI is understood as consisting purely of software resources–such as data, algorithms, and machine learning models–then AI developers, users, and regulators will be bounded within a very different governance system than if AI were interpreted as also consisting of, for example, the human labor and knowledge, hardware components, digital infrastructure, energy, water, and minerals required to develop and operate software systems.

Expanding the ontological perimeter of AI to encircle a broader range of phenomena supports more interpretive frames, discursive spaces, values and norms, and practical possibilities for AI governance. By allowing ambiguity and fluidity in how the term AI is perceived, interpreted, discussed, and acted upon, the innate *transness* of AI can function as a foundation for more inclusive forms of AI governance. In a 2023 essay, transfeminist scholars Mijke van der Drift and Nat Raha write that "radical transfeminism embraces trans as active and anti-normative, rather than defined as a stable form . . . Trans is thus a dynamic formation, which does not lay claim to simply *be*, but which functions by disrupting static categories of being" (pp. 13-16). At a time when so many new laws, policy initiatives, and technical standards are attempting to stabilize AI into a well-defined form (see for example the definitions of AI that form the basis of the EU's *AI Act* and the ISO/IEC 42001:2023 standard for AI management systems), it is imperative that a transfeminist approach to AI governance resists forced stabilization into narrow techno-legal definitions and forms of governance. A transfeminist approach to AI governance can instead embrace the transness of AI: the ambiguity, the fluidity, and the plurality of meaning embedded in the term AI all provide affordances for more diverse understandings of AI, epistemic communities, governance systems, and justice-seeking projects to form within the vast discursive nebula we refer to as "AI."

Like AI, the concept of "AI governance" also has some innate transness. AI governance is a dynamic and ambiguous phenomenon, simultaneously referring to many possible ways of researching, practicing, and systematizing interventions in "AI." In this article, some disambiguation has been provided by framing AI governance as a system of practices intended to maximize benefits and minimize harms caused by AI systems. However, this framing is itself deliberately ambiguous, intended to transform AI governance from a domain dominated by legal, technical, business, and policy experts into a more open discursive space that can accommodate diverse perspectives, exploratory questions, and alternative answers. What types of benefits are included in AI governance, and why? What types of harms are included, and why? Who benefits and who is harmed, and through what types of practices do they intervene in AI systems to maximize benefits and minimize harms? Studies 1, 2, and 3 all provide some potential answers to those questions, but the amorphous ontological perimeter of AI and the fluid contexts, scales, and scopes of AI governance systems leave space for many alternative answers. There is no definitive, objective answer to what AI governance is and what the work of AI governance ought to entail. Ontologies and ethics of AI governance are co-created by many networks of actors applying their many resources, logics, rules, and norms with the intent of planning and enacting governance activity. In Study 3, we saw that AI governance contains a plurality of values, perspectives, and practices. It could be said that writing and reading this article are both practices of AI governance: we write and read the research presented here with the intent of cultivating knowledge that can be applied to maximize benefits and minimize harms caused by AI systems.

AI governance is not the preserve of well-funded experts, academics, lawyers, and public intellectuals. A transfeminist approach to AI governance resists what van der Drift and Raha (2023) refer to as "neoliberal encapsulation:" the containment of transness to "the future that liberalism has wanted for us – a future based in limited forms of social inclusion and legal rights" (p. 13). A transfeminist approach to AI governance works against neoliberal encapsulation by expanding the ontological scope of AI governance to include a more comprehensive set of actors and activities than what is typically found within the narrow technocratic confines of state-led and industry-led AI governance initiatives. Communities and workers impacted by AI systems all possess their own domain expertise: they possess context-sensitive, direct knowledge of AI impacts and resource needs within their communities and workplaces. Collective organizing, community-led policy and strategy, direct action against harmful systems and policies, development of small-scale resource-sharing and capacity-building networks, discursive interventions, and public protest can all be directed toward maximizing benefits and preventing harms caused by AI systems. Therefore, these are all legitimate practices of AI governance. By expanding the perimeter of AI governance to encompass this larger scope of practical possibilities, AI governance is transformed into a more inclusive democratic project, one that escapes neoliberal encapsulation. Rather than consolidating more expertise, agency, and power into already powerful state and industry organizations, a transfeminist approach to AI governance must build greater agency and power within marginalized and under-resourced communities that have so far had their needs largely ignored by state and industry actors.

**Action 2: Oppose AI governance systems that exclude marginalized communities.** As it has become more clear that AI systems frequently cause harm, a vibrant discussion has emerged regarding the accountability processes through which impacted actors may oppose AI developers and operators whose systems caused harm to them. Principles of public accountability, contestability, and resistance are of particular importance in these discussions, and conceptual frameworks such as "algorithmic resistance" (Bonini and Treré, 2024; DeVrio et al., 2024; Ganesh and Moss, 2022; Velkova and Kaun, 2021), "resisting AI" (McQuillan, 2022) and "contestable AI" (Alfrink et al., 2023; Alfrink et al., 2024; Balayn et al., 2024) have emerged as banners under which diverse practical possibilities for opposition against AI systems are studied. Although these frameworks share a common interest in studying methods through which the developers and operators of AI systems can be opposed by vulnerable and harmed stakeholders, contestability-based frameworks and resistance-based frameworks do have some crucial differences and limitations.

Alfrink et al. (2023) characterize *contestability in AI* as distinctively procedural and mechanistic. Contestability requires institutional mechanisms for open and iterative debate about AI design practices, responsiveness to complaints about an automated decision or other system output, and commitments to remediation and procedural justice throughout the lifecycle of the system. Although contestable AI provides a process-oriented approach for opposing context-specific harms caused by AI systems, the effectiveness of contestability mechanisms is largely dependent upon voluntary commitments to participatory design and procedural justice from AI developers, operators, and policymakers. Many of these actors would, again, be acting against their own perceived interests by empowering marginalized actors with robust mechanisms with which to contest their practices of AI development and policymaking. Indeed, Alfrink et al. imply that contestable AI faces complex structural barriers to democratic control even in public sector contexts where it is more likely to be effective, noting the "duty of care government organizations have toward citizens; and the (*at least nominal*) democratic control of citizens over public organizations" (p. 632, emphasis added).

Meanwhile, *resistance against AI* refers to a broader category of oppositional acts that emerge from structural conflict with power within an AI system, including (but not limited to) acts of contestation and complaint. In their review of the research literature on algorithmic resistance, Bonini and Treré (2024) observe that resistance against AI systems and other data-intensive algorithmic systems entails an *agonistic* relationship with power, a type of structural relation between dominant and marginalized actors that is simultaneously both frictive and generative of new possibilities for building greater agency amongst marginalized actors. Bonini and Treré are careful to diminish the revolutionary romanticism of resistance, noting that although resistance is "born of a response to existing systems of power, control, and domination," (pp.17-18), acts of resistance are not limited to historically marginalized actors: resistance is practiced by a range of actors seeking to cultivate greater agency that they can then exercise to enact a variety of benevolent and malevolent intentions.

The limitations of these frameworks led me to write an essay entitled *AI Countergovernance* intended for a wide audience of researchers, practitioners, public servants, community organizers, and activists seeking strategies and tools for opposing harmful AI systems (Attard-Frost, 2023c). In the essay, I apply transfeminist principles of fluidity, agency, community, resistance, and transformative change to outline a framework of oppositional action that reconciles the mechanistic view of contestability-based frameworks with the structural view of resistance-based frameworks. I build upon Dean's (2018) analysis of contestability mechanisms and resistance practices involved in building counterpower against state actors whose governance systems failed to serve the needs of their publics. Dean refers to this process as "counter-governance." I introduce *AI countergovernance* as a conceptual and practical framework for opposing AI governance systems that exclude the needs of marginalized communities. Synthesizing lessons learned from four real-world cases of collective organizing against state-led and industry-led AI governance systems (see Table 3), I argue that communities and workers can more effectively oppose harmful AI systems by organizing against their higher-level governance systems (e.g., by organizing against the full spectrum of organizational actors, networks, logics, and resources responsible for co-creating the harmful AI system), rather than by centering contestation and resistance around more granular technological components and socio-technical practices involved in developing and operating those components (e.g., the development and use of harmful datasets, algorithms, and machine learning models.)

*Table 3: Key lessons learned from the four cases of AI countergovernance discussed by Attard-Frost (2023c), highlighting comparative sources of power and counterpower, practices of countergovernance, and outcomes of each case.*

| Case | Sources of power & counterpower | Practices of countergovernance | Outcomes |
|---|---|---|---|
| Project Maven (Google military partnership with US Department of Defense) | • *Power*: Industry-state partnership<br>• *Counterpower*: Workers | • Open letter to Google CEO<br>• Worker resignations & walkouts<br>• Media engagement | • Project Maven contract was not renewed in 2019<br>• Re-emergence & expansion of Google-DoD partnerships in subsequent years |
| Sidewalk Toronto ("Smart city" development project) | • *Power*: Industry-state partnership<br>• *Counterpower*: Local community | • Public meetings<br>• Alternative urban development & data infrastructure plans<br>• Media engagement | • Sustained community backlash against project<br>• Cancellation of project in May 2020 due to multiple factors (including resistance from local community) |
| Backlash against Canada's Artificial Intelligence & Data Act (AIDA) | • *Power:* State with significant industry influence<br>• *Counterpower:* Public (Primarily policy experts & advocacy groups) | • Open letters to Members of Parliament<br>• Public audits & assessment of the AIDA<br>• Alternative policy proposals<br>• Witness testimony & impact statements<br>• Media engagement | • Sustained public backlash against AIDA<br>• New consultations launched & amendments to AIDA that reflect some criticisms<br>• Many criticisms currently remain unaddressed |
| Writers Guild of America Strike | • *Power*: Industry<br>• *Counterpower*: Workers | • Collective bargaining<br>• Policy & planning through workplace technology committee<br>• Media engagement | • New collective agreement contains regulations for AI-generated writing, writer use of AI, and use of writers' material in model training |

This countergovernance framework first expands the ontological scope of "AI" and "AI governance" to accommodate a more open horizon of practical possibilities, then moves the target of opposition from AI systems to AI governance systems. In this framework, communities organize against ineffective AI governance systems by working transversally both inside and outside of established institutions, as well as by leveraging a fluid repertoire of governance practices to build agency and counterpower. Practices of AI countergovernance include awareness-building through media engagement, protests, petitions, and open letters addressed to sources of state and industry power, such as elected representatives and corporate executives. Countergovernance can also include community-led audits, evaluations, and iterative contestation of harmful AI systems, policy instruments, and organizations. Successful countergovernance processes can result in co-creation of shared knowledge resources and funds, guidelines and rules, and prevention of the development and operation of harmful AI systems within a community. Through its grounding in transfeminist principles and practices of governance, this framework enables actionable strategies of contestation, resistance, capacity-building, and justice-seeking within marginalized communities that are vulnerable to or have been harmed by AI systems.

**Action 3: Co-create bottom-up AI governance systems to reduce dependency on industry and the state.** Beyond merely opposing ineffective large-scale systems of top-down AI governance, a transfeminist approach to AI governance can also advance the bottom-up development of small-scale alternative systems in more localized contexts, such as communities and workplaces. If developed in accordance with transfeminist principles of transformative change, agency, and community, a foundational goal of developing bottom-up AI governance systems should be to gradually reduce the dependence of vulnerable communities on AI governance resources provided by state and industry actors. Transfeminist AI governance must build within communities the resources, networks, and agency needed to exercise greater power over AI systems that pose a risk to the community. As governance practices, actors, networks, resources, logics, rules, and norms within the community gradually become more systematized, power can be more effectively built and exercised to prevent the development and operation of AI systems that may cause harm to the community. This will be a fluid and dynamic process, involving diverse and extensive co-creation activity between community organizers and researchers, practitioners, public servants, journalists, activists, and other sympathetic actors with the intent of supporting the community in building and exercising power.

**Researchers** can support community organizers with empirical research and action research intended to determine more clear, context-sensitive practices for effective community-led design and implementation of AI governance systems. **Practitioners** of technology design, management, and policy can support by training community organizers with skills and knowledge from their respective disciplines. Practitioners can assist community and labor organizers in designing and implementing projects and programs, policy instruments and guidance, community events and workshops, knowledge resources, technologies, and other governance tools. **Public servants** can support by directing a greater share of public funding and resources away from industry-centric AI governance initiatives and toward community-centric governance initiatives. These can include knowledge-building initiatives and participatory policymaking in historically marginalized communities, as well as civic participation funds to enable under-resourced civil society organizations and communities to more actively participate in policy design and co-regulation. Resources can also be directed toward creating legal supports and working groups to build stronger regulatory infrastructure and capacities for resistance. **Journalists and creatives** can support by investigating the impacts of AI on the community, giving voice to the experiences and stories of the community, and building public awareness of the realities, possibilities, and potential futures of community-led AI governance. This work is particularly crucial for helping to build the cultural power and counternarratives needed to oppose dominant futurological narratives of AI-fueled utopias, economic windfalls, and technocratic regulators staving off the existential risk of unaccountable "rogue AI" systems. Unaccountable AI systems already exist in the present moment; their developers and operators are already self-regulating rogues who pose an existential risk to marginalized communities.

Counternarratives such as this are essential for subverting the discursive frames and regulatory pretexts of dominant actors.

## 6. Conclusion

A transfeminist approach to AI governance is rooted in the lived experiences of trans people, and the histories of domination, marginalization, and oppression that trans people–especially Black trans people and trans people of color–have been forced to endure. This article has emphasized the application of transfeminist ethical principles to AI governance research and practice over more direct engagement with the lived experience of being impacted by an AI system as a trans person. However, a recent report by a group of researchers at the University of Virginia shows that AI and other data-intensive technologies are indeed causing trans communities around the world to experience numerous physical, psychological, social, political, and economic harms (Reia et al., 2025). Against a backdrop of rising anti-trans sentiment and anti-trans legislation around the world, their report underscores an urgent need to develop community-centric systems of AI, data, and digital governance that protect trans people and empower us to prosper in online and offline environments. As Susan Stryker (2017) shows in her pivotal work *Transgender History: The Roots of Today's Revolution*, the history of trans liberation movements provides ample evidence that trans communities cannot fully rely upon or entrust the governance systems of the state to protect us from harm. Not only are state-led systems of governance frequently developed without meaningful inclusion of trans and other marginalized communities, but in far too many instances, *state power is actively exercised with the intent of causing harm to us*. It is only through arduous acts of community organizing, direct action, and bottom-up practices of governance involving fluid, decentralized networks of collaborators–both inside and outside the halls of institutional power–that some political victories have been won for some trans communities in some parts of the world.

Future AI governance research, practices, and systems have much to learn from trans ways of thinking, feeling, and doing governance. It is not sufficient for us to do nothing more than wait for alliances of state and industry power to reach accords on legal frameworks for AI, even if we are granted a seat at the table of their discussions. It is not sufficient for us to hope that those laws will be effectively and justly enforced to protect trans lives, especially in the present context of intensifying anti-trans violence led and sanctioned by state actors. A transfeminist future for AI governance begins within our communities and outside of the state. As Stryker puts it in concluding her account of the history of trans liberation movements (2017, p. 236): "We can do more than cross our fingers and hope for the best if we ourselves work together to bend our little corner of the universe."

**Acknowledgements**

This article has greatly benefited from the generous guidance and feedback of my doctoral supervisory committee members Kelly Lyons, Leslie Shade, and Chun Wei Choo, and from countless conversations with my friends and collaborators Ana Brandusescu and David Gray Widder. A very special thank you to the many friends and family who've supported me throughout this research. This article draws on research supported by the Social Sciences and Humanities Research Council of Canada (SSHRC).